\documentclass[%
 aip,
 amsmath,amssymb,
 reprint,%
]{revtex4-2}

\usepackage{graphicx}
\usepackage{dcolumn}
\usepackage{bm}
\usepackage{float}
\usepackage{ulem}
\usepackage{color,soul}
\usepackage{wrapfig}
\usepackage{graphicx}
\usepackage{amsmath}
\usepackage{amsfonts}
\usepackage{latexsym}
\usepackage{amsbsy}

\usepackage{bm}%
\usepackage{subfig}
\usepackage[utf8]{inputenc}
\usepackage[T1]{fontenc}
\usepackage{mathptmx}
\usepackage{etoolbox}
\draft 

\begin{document}


\title{TV and Video Game Streaming with a Quantum Receiver: A Study on a Rydberg atom-based receiver's bandwidth and reception clarity
} 
\thanks{Publication of the U.S. government, not subject to U.S. copyright.}


\author{Nikunjkumar Prajapati}
\author{Andrew P. Rotunno}
\author{Samuel Berweger}
\author{Matthew T. Simons}
\author{Alexandra B. Artusio-Glimpse}
\author{Christopher L. Holloway}
\affiliation{{National Institute of Standards and Technology, Boulder,~CO~80305, USA}}


\date{\today}

\begin{abstract}
We demonstrate the ability to receive live color analog television and video game signals with the use of the Rydberg atom receiver.
The typical signal expected for traditional 480i NSTC format video signals requires a bandwidth of over 3~MHz.
We determine the beam sizes, powers, and detection method required for the Rydberg atoms to receive this type of signal. 
The beam size affects the average time the atoms remain in the interaction volume, which is inversely proportional to the bandwidth of the receiver.
We find that small beam diameters (less than 100 $\mu$m) lead to much faster responses and allow for color reception.
We demonstrate the effect of beam size on bandwidth by receiving a live 480i video stream with the Rydberg atom receiver. The best fidelity was achieved with a beam width of $85~\mu$m full-width at half-max.
\end{abstract}


\maketitle 

\section{Introduction}

Rydberg states (highly excited) of atoms have been of growing interest in the past decade and have provided an avenue for making a variety of different sensors~\cite{9748947}.
This is possible since Rydberg states are highly sensitive to electric fields and depending on the Rydberg state used, allow for detecting fields ranging from DC to THz.
They have been used for the detection of electric fields for AM/FM (and phase modulation) receivers~\cite{Song:19, doi:10.1063/1.5028357, Cox_2018,9069423, 9054945,8778739,  rydberg_array, doi:10.1063/1.5099036}, spectrum analyzers~\cite{PhysRevApplied.15.014053}, voltage standards~\cite{plate_cell_sandia}, angle-of-arrival~\cite{aoa}, and many more applications~\cite{Jing2020,Simons2019ApplicationsWA,Norrgard_2021,Repump_paper,SRR}. 
These sensors even allow for calibrated measurements traceable to the international system of units (SI)~\cite{SI_standard} for both electrical and radio frequency power~\cite{holl1, pow_stand, Sedlacek_2013, si_trace}. 

A current thrust in the development of Rydberg atom-based sensors is geared towards improving the sensitivity and in understanding the limits of bandwidth for these quantum-based receivers. 
While the reception of various signals have been shown, the reception of live television (TV) has yet to be demonstrated.
This limitation largely arises form the bandwidth to signal relationship.
A recent study demonstrated this link between the bandwidth and signal strength~\cite{beam_size_shaffer}.
We take this a step further to optimize the signal/bandwidth and utilized optical homodyne detection to receive a TV/gaming signal that can be directly output from a photodetector to a TV.
\begin{figure}[h!]
    \centering    
    \includegraphics[width=\columnwidth]{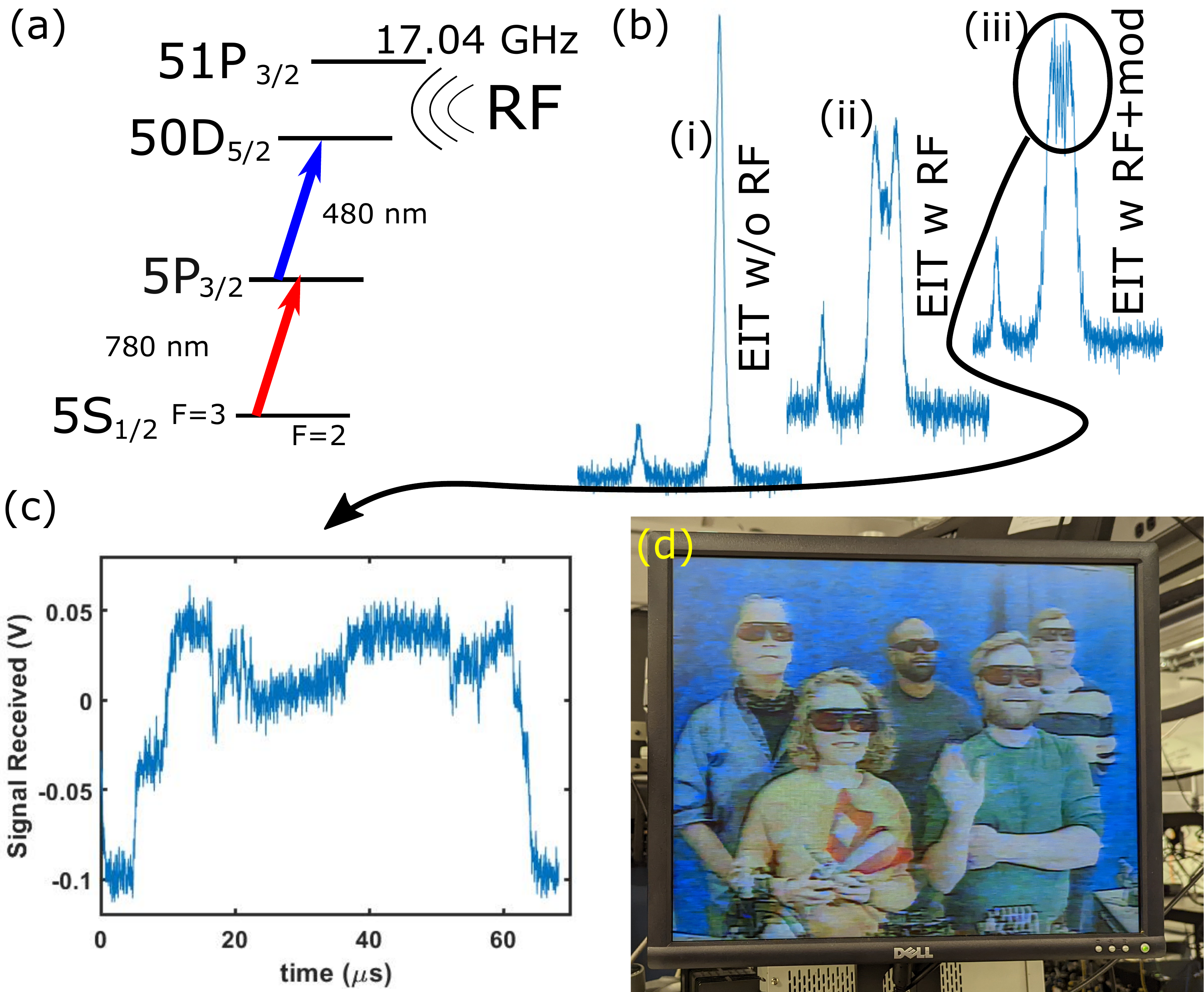}
    \caption{(a) Level diagram depicting EIT coupling the 5S$_{1/2}$ and 50D$_{5/2}$ states through the 5P$_{3/2}$ intermediate state. Radio frequency (RF) field couples the Rydberg states 50D$_{5/2}$ and 51P$_{3/2}$
    (b) Sample EIT resonance (i) without RF carrier, (ii) with RF carrier, and (iii) with RF carrier being modulated. (c) Modulation present on one row of the TV signal showing the trigger and pixels. (d) Live TV image obtained through output from the photo-detector with no filtering or external amplification (see video file attached for live feed).}
    \label{fig:level_tv}
\end{figure}

In this demonstration, we utilized electromagnetically induced transparency (EIT) to probe the Rydberg state of interest ($50D_{5/2}$), shown by Fig.~\ref{fig:level_tv} (a) and trace in (b). Then, we apply a radio frequency (RF) field that is resonant with the $50D_{5/2}\rightarrow 51P_{3/2}$ transition which results in the Autler-Townes (AT) splitting of the Rydberg state, as shown by another trace in Fig.~\ref{fig:level_tv} (b). This 17.0434 GHz field is the carrier of our signal. The signal is in the form of an analog amplitude modulation of the carrier, shown by Fig.~\ref{fig:level_tv} (c). By locking the laser to the center of the EIT resonance, the energy level shift translates to an amplitude modulation of the transmission signal of the laser~\cite{Song:19, 9054945, 9069423}. This modulation of the laser is then detected by the photo-detector and can be fed directly to a CRT TV. Here, we used an analog to digital converter to transform the analog signal into video graphics array (VGA) format to display on a monitor, shown by Fig.~\ref{fig:level_tv}(d).

\begin{figure}[h!]
    \centering
    \includegraphics[width=\columnwidth]{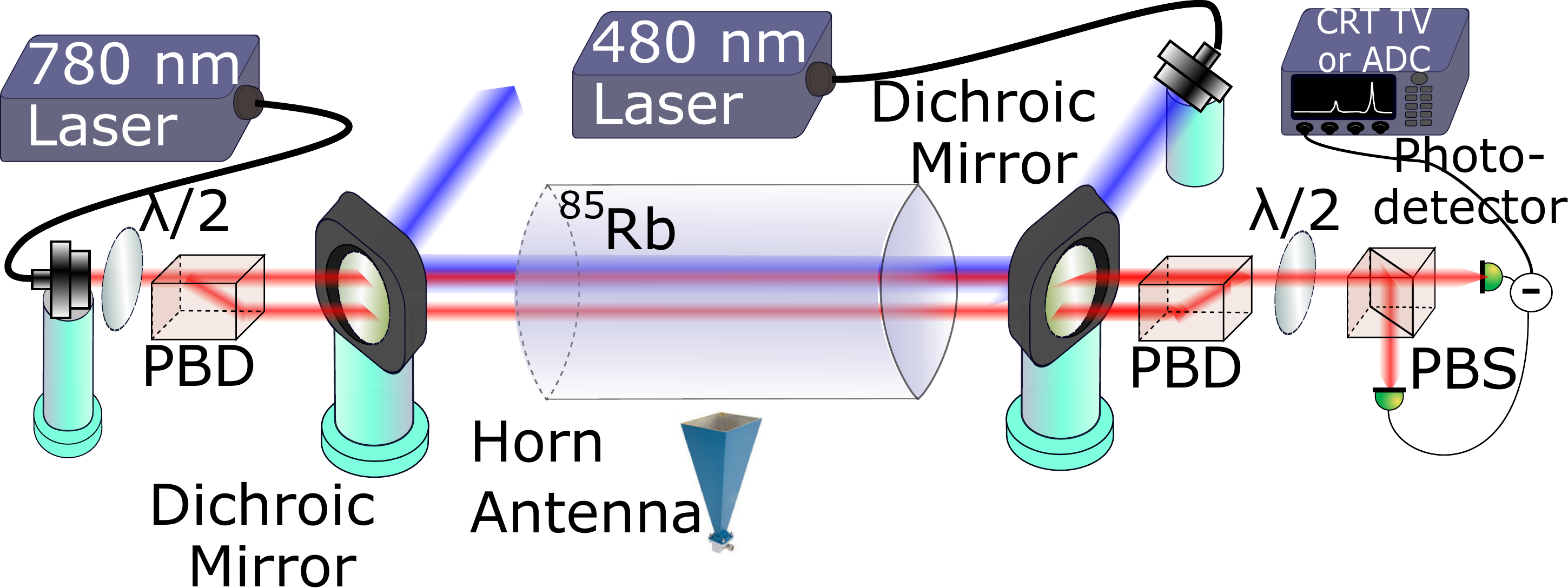}
    \caption{Schematic showing the experimental setup.}
    \label{fig:schematic}
\end{figure}

In this manuscript, we present a detailed study on how the signal and bandwidth depend on the beam sizes and powers and how these conditions affect the reception of the live video feed, which ultimately determines the clarity of the reception and if color can be received.
Fig.~\ref{fig:level_tv} (d) shows a clean reception of a TV signal, with optimal beam size and optical power.
In this experiment, we used the setup shown in Fig.~\ref{fig:schematic}. 
This setup allowed us to switch between homodyne detection and balanced differential detection depending on the observed signal.

\section{Experimental Setup}

We use a 780~nm external-cavity diode laser (ECDL) as our probe laser and a 960~nm ECDL amplified by a tapered amplifier that is fed into a second harmonic generation cavity to generate our 480~nm coupling laser. Both lasers' powers at the cell are stabilized at the cell location by using acoustic optic modulators. The coupling laser power is fixed to 70~mW for this study.

The polarizing beam displacer (PBD) splits the 780~nm laser into a signal and reference beam. The signal beam is overlapped with the coupling laser in the vapor cell while the reference beam is not. After interaction in the cell, both signal and reference beams are overlapped spatially using the second PBD. The waveplate slow axis is set to either 0$^\circ$ or 45$^\circ$. When we want to utilize balanced differential detection, the waveplate is set to balance the signal and reference power and the second waveplate is set to be aligned with the slow axis. 
For homodyne detection, the first waveplate is set so that the signal beam power is 30~$\mu W$ and the reference beam is $\sim 1.5$~mW. We stabilized the system with boxes to reduce fluctuations from air currents.

We utilized differential detection when taking power dependent data for long term stability and used homodyne detection for added signal strength at lower detector gain. The detector used in these experiments has limited bandwidth at high gain, so we employed homodyne detection to increase the signal such that we could operate the photodetector with lower gain~\cite{homodyne_vs_balanced_detection}.

\begin{figure}[h]
    \centering
    \includegraphics[width=\columnwidth]{"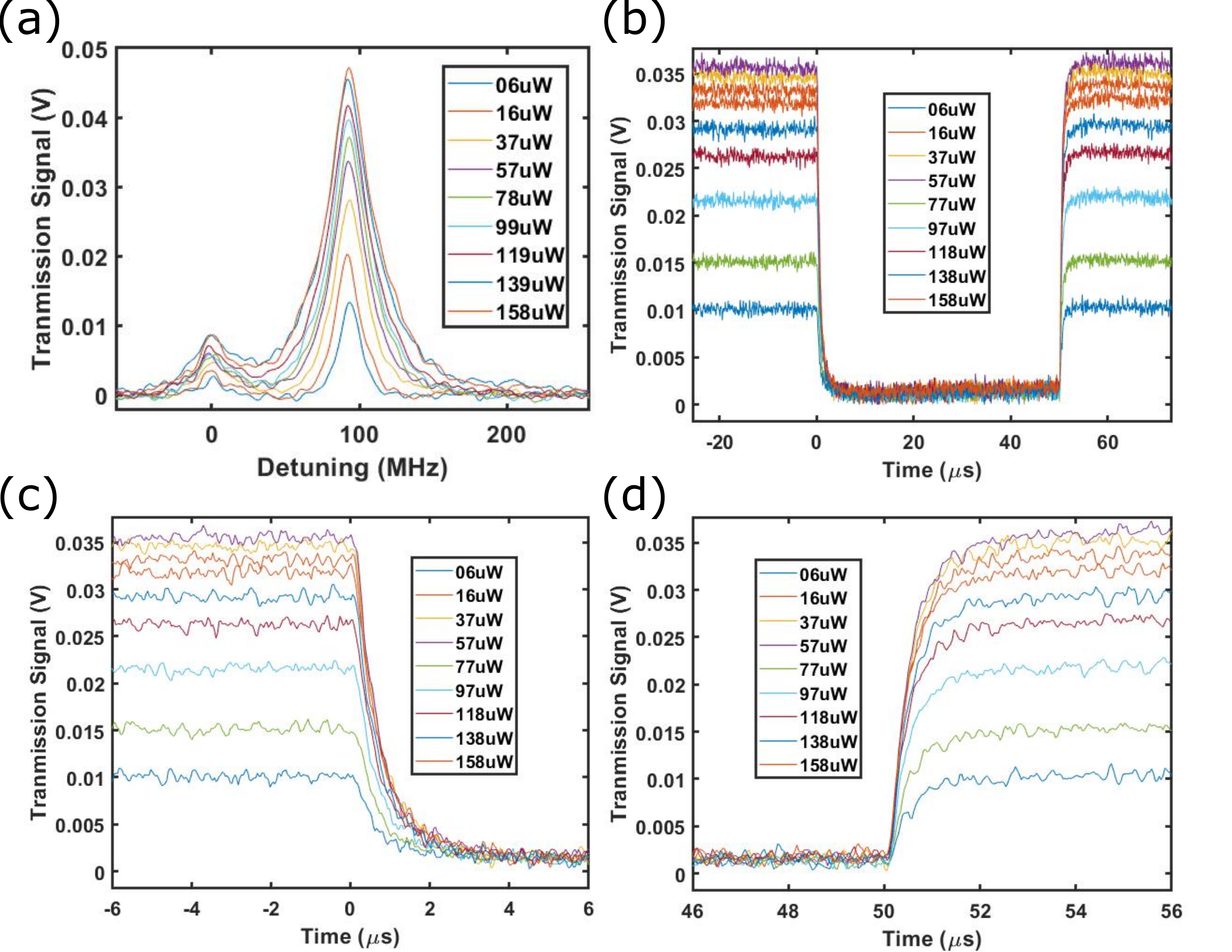"}
    \caption{(a) EIT resonance as the coupling laser is scanned for different probe powers. (b) Squarewave detected when coupling laser is locked to the EIT peak and the RF source is modulated at 10 kHz (for different probe powers). (c) Zoom in on the fall time of (b). (d) Zoom in on the rise time of (b). This data is for a beam size of 85 $\mathrm{\mu m}$}
    \label{fig:EIT_and_square}
\end{figure}

For these experiments, we varied the probe laser beam size and adjusted the coupling laser beam width to keep it slightly larger than the probe. This avoided unwanted effects from spatial intensity variations. The beams were focused at the center of the cell to achieve the desired beam waist.
For each beam width we varied the power of each beam to change the Rabi frequency. Fig.~\ref{fig:EIT_and_square}(a) shows the EIT signal as a function of probe Rabi frequency for a beam full-width at half-maximum (FWHM) of $85~\mu$m. We also measured the rise and fall times of the atomic response by locking both laser frequencies to the EIT resonance and applying a $10~$kHz square wave. This was done by locking both the probe and coupling laser to the EIT resonance and observing how the RF field effected the transmission signal. The 17.0434 GHz RF field was generated by a signal generator and was mixed with a square wave modulation from a function generator using a RF mixer. This modulated RF field was then fed to a horn antenna to radiate the atoms, shown by Fig.~\ref{fig:EIT_and_square}(b-d). 

\section{Experimental Results}
\begin{figure}[h!]
\centering
    \includegraphics[width=\columnwidth]{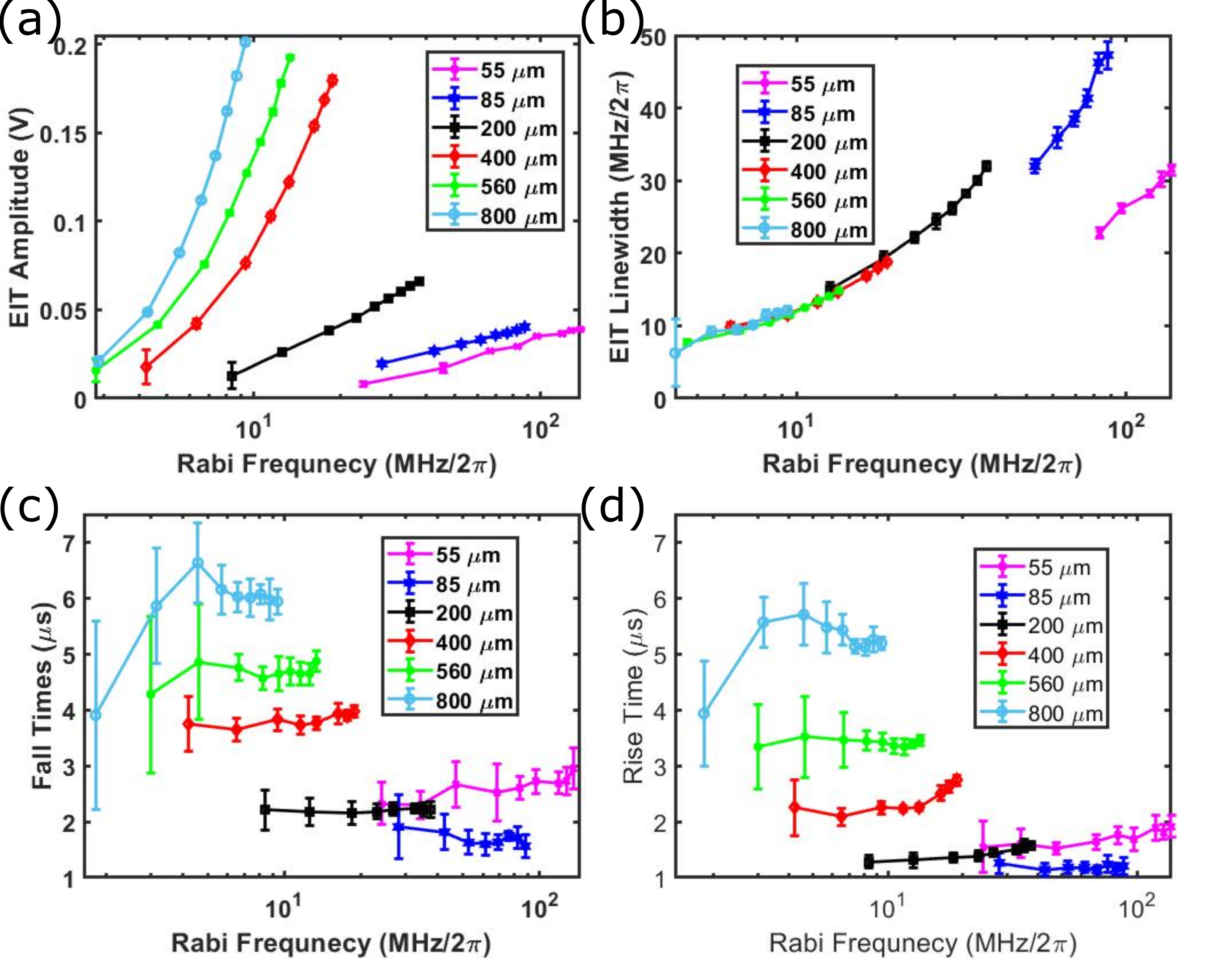}
    \caption{(a) EIT FWHM, (b) EIT amplitude, (c) fall time, and (d) rise time plotted against the probe Rabi frequency (measured at beam waist) for the different beam sizes. This data was extracted from the data similar to that shown in Fig.~\ref{fig:EIT_and_square}}
    \label{fig:widths_and_heights}
\end{figure}

To demonstrate the effects of beam size and powers on the EIT signal and response times, we collect traces, similar to those shown in Fig.~\ref{fig:EIT_and_square} for different probe beam powers and beam waists.
From these measurements, we extract the EIT height, EIT width, rise times, and fall times, shown by Figs.~\ref{fig:widths_and_heights}.  Fig.~\ref{fig:widths_and_heights}(a) shows the height of the EIT peak as a function of probe laser Rabi frequency for each beam width. The height of the EIT peak is proportional to the number of atoms that take part in the EIT interaction, which is proportional to the interaction volume. Increasing beam width increases the interaction volume, so for the same Rabi frequency a larger beam results in a stronger EIT peak. Figs.~\ref{fig:widths_and_heights}(b) shows the EIT linewidth as a function of probe Rabi frequency for each beam width. As opposed to the height, the linewidth does not depend on the beam size and is proportional to the Rabi frequency of the probe laser. 
Note that the measurements for a $55~\mu$m beam width is not in line with the rest of the traces. This is due to the strong divergence of the probe beam used to obtain the tight beam waist at the center of the cell. 
The Raleigh range for this beam is 24 mm smaller than the length of the cell.
The beam width changed substantially through the cell, so the effective Rabi frequency is much lower than estimated. The other beam widths did not diverge substantially through the cell. 

We next investigated the rise and fall times for the atomic response to a square wave as shown in  Fig.~\ref{fig:widths_and_heights}(c-d). 
The temporal response does not significantly depend on the probe Rabi frequency, as opposed to the EIT height. However, both the rise and fall times depend strongly on the beam width.
These times are proportional to the beam width, which changes the length of time an atom spends in the interaction volume. 
Since atoms are moving in and out of the interaction volume, smaller interaction volumes have a larger `refresh rate' for the interaction. 
The average velocity of room temperature Rubidium atoms is roughly 240~m/s. So an atom will be able to transit through the interaction region within the transit time $t_{transit}$ as given by
\begin{equation}
    t_{transit} = \frac{2\cdot\omega}{\nu}
    \label{eq:transit}
\end{equation}
where $\omega$ is the beam waist (given by FWHM$/\sqrt{2ln(2)}$) and $\nu$ is the average atomic velocity. We compare the average rise time and fall times to the average transit times for the different beam size, shown by Table.~\ref{tab:rise_fall_times}.

\begin{table}[h]
    \centering
    \begin{tabular}{c|c|c|c|c}
    Probe  & Coupling  & Rise  & Fall  & Transit  \\ 
    FWHM ($\mu$m) & FWHM ($\mu$m) &  Time ($\mu$s) & Time ($\mu$s) & Time ($\mu$s)\\ \hline
    55  & 120 & 1.5 & 2.6 & 0.38 \\ \hline
    85  & 120 & 1 & 1.2 & 0.60  \\ \hline
    200 & 220 & 1.2 & 2.1 & 1.41 \\ \hline
    400 & 800 & 2.1 & 3.8 & 2.83 \\ \hline
    560 & 800 & 3.2 & 4.5 & 3.96 \\ \hline
    800 & 800 & 5.4 & 6.1 & 5.66
    \end{tabular}
    \caption{Table showing the rise times, fall times, and transit times for the different beam widths. The rise and fall times were extracted from the data and the transit time is found using Eq.~\ref{eq:transit}}
    \label{tab:rise_fall_times}
\end{table}

We found that the rise and fall times vary with the transit time, except for the $55~\mu$m width. Again, this beam has a much higher divergence, so a significant proportion of the beam is much larger than the beam waist.
In addition to this, we note that for some cases that the rise time is faster than the transit time. This is due to the rise time being independent of the atomic decay. Establishing a Rydberg population is dependant on the effective Rabi strength of the two-photon interaction.

These results allowed us to optimize the beam size for bandwidth, and achieve a sufficiently fast response to stream live video signals from a video camera and from a video game console. For this demonstration, we used an optical homodyne setup, with a signal beam of 30~$\mu$W and reference beam of $1.5~$mW. The half-waveplate near the photodetector in Fig.~\ref{fig:schematic} rotated the polarization of the signal and reference beams by $45^\circ$ with respect to the polarizing beam cube slow-axis to mix the reference and signal on the photodetector~\cite{Kumar2017}.

\begin{figure}[h!]
    \centering
    \includegraphics[width=\columnwidth]{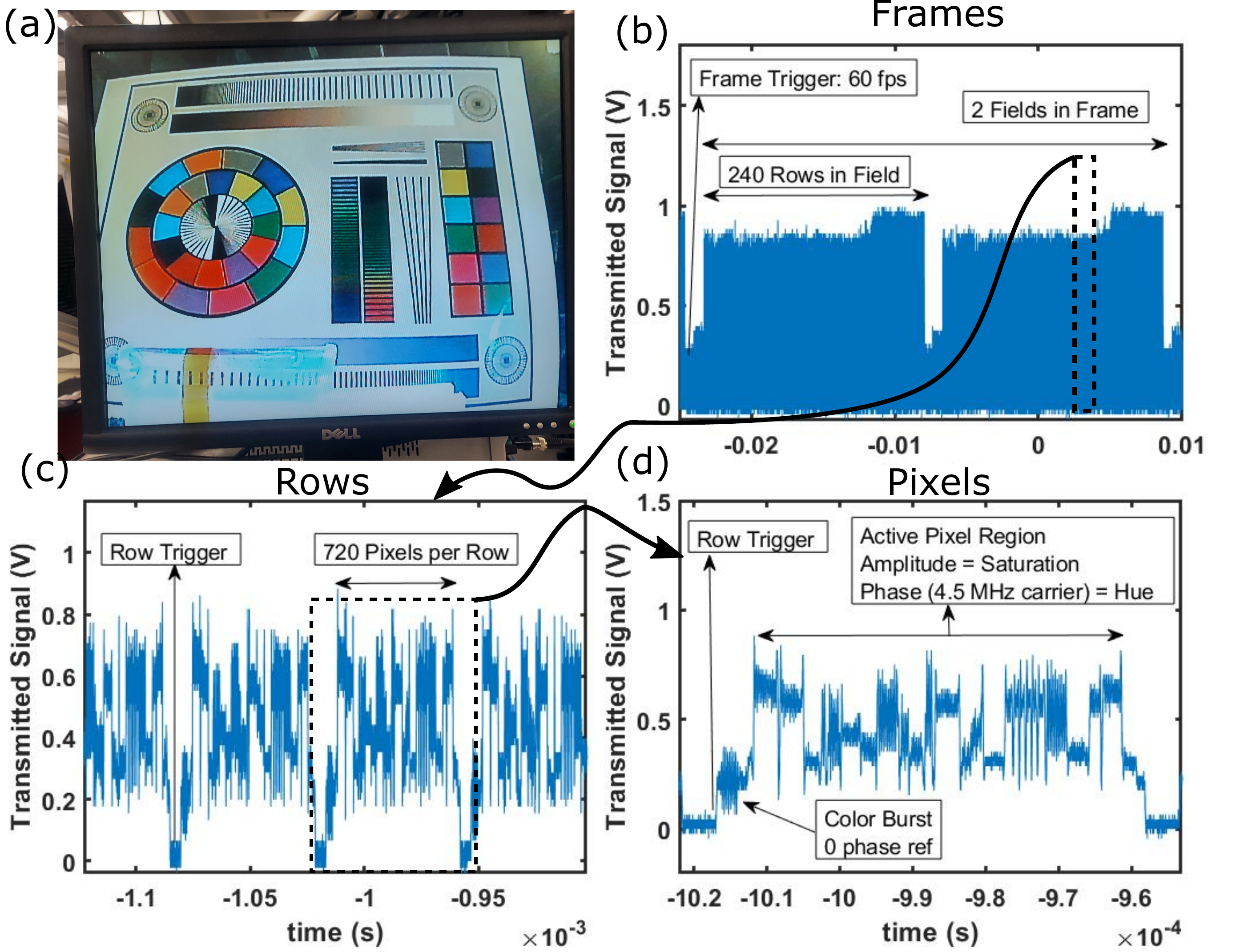}
    \caption{The description of the NTSC 480i video format shown by direct output from video camera to oscilloscope: (a) Image of direct signal to the analog-digital convertor, (b) Larger time span that captures several frames and their trigger, (c) Zoom in on Frame that captures three rows and their triggers, and (d) Zoom in on row that captures information for each pixel and the color burst that defines the phase. In NTSC 480i, a $3.58~$MHz wave carries the pixel saturation in amplitude and color in phase.}
    \label{fig:NTCS_disc}
\end{figure}

\begin{figure*}[t!]
    \centering
    \includegraphics[width=\textwidth]{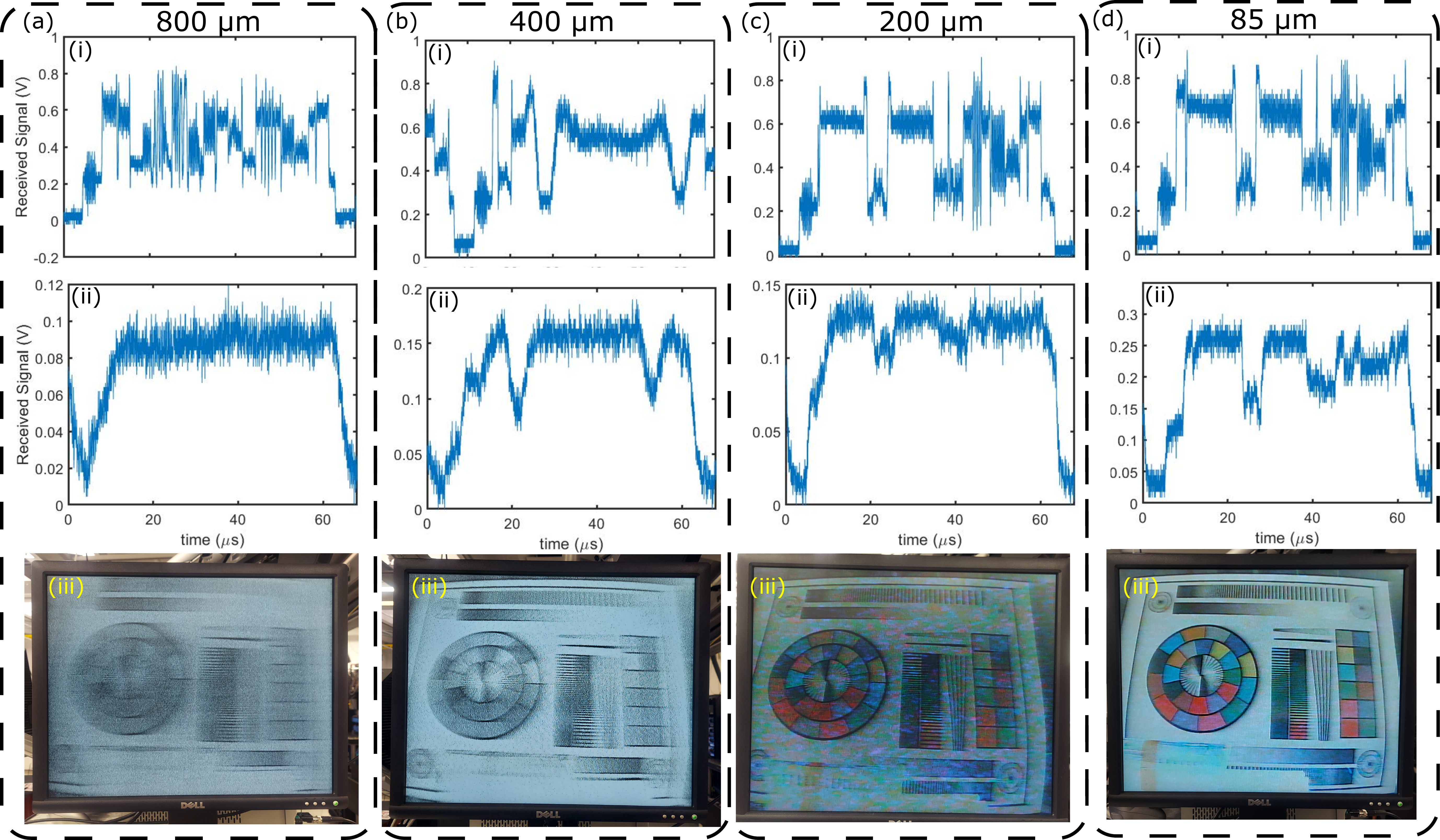}
    \caption{(a-d) Columns show sent and received data from camera for different beam sizes as labeled. (i) Live video signal from video camera for a given row in NTSC video format. (ii) Signal received by the atoms for the same rows as (i). (iii) Video received for the different beam sizes for the given column.}
    \label{fig:Lines_and_rates}
\end{figure*}

The video format that is output by the camera and the video game console used in this study was NSTC 480i, or standard definition. In this example we streamed the video of a printed color test pattern shown in Fig.~\ref{fig:NTCS_disc}(a). The direct signal output of the 
video camera was shown by Fig.~\ref{fig:NTCS_disc}(b)-(d).  The signal is an analog waveform that gives information on the frames, rows, and pixels. Fig.~\ref{fig:NTCS_disc} (a) shows several fields (each is half of an interlaced frame), where we have identified the trigger marker for a given one. We also label the 240 active rows for each field.  Fig.~\ref{fig:NTCS_disc} (b) shows several rows for a given field and the trigger for each row. Fig.~\ref{fig:NTCS_disc} (c) gives the information for each row. After each row trigger, there is a `colorburst' signal that sets the reference phase for the 3.58~MHz carrier that determines the color. Each pixel in the active area of the row is represented by one quarter-cycle of 3.58~MHz carrier, yielding roughly 720 pixels per row. The amplitude of the cycle gives the saturation of color, and the phase of the cycle relative to the colorburst gives `chrominance' or hue color information. The color depth for the video camera is nominally 24-bit, with 8 bits of information in each of the red, green, and blue basis colors.
Brightness or `luminance' is given by the signal offset, making this format backwards compatible with black-and-white images when the color information is not present, as seen in Fig.~\ref{fig:Lines_and_rates}. 

To receive the video signal, both lasers were locked to the EIT resonance. A local oscillator RF carrier was used to set the AT splitting to the highest sensitivity region. The signal from the video camera modulated a $17.04~$GHz carrier using an RF mixer. The modulated carrier was fed to a horn antenna to direct the field to the atoms. We compare the original video signal to the down-converted signal on the probe laser received through the atoms in Fig.~\ref{fig:Lines_and_rates}(i-ii) for each column (a-d). The columns show the measurement results for three different beam sizes, $800~\mu$m, $400~\mu$m, $200~\mu$m, and $85~\mu$m. The last row in each column (iii) shows the demodulated video displayed on a screen. 
We see that for the larger beam size (800~$\mu$m FWHM), the received signal Fig.~\ref{fig:Lines_and_rates}(a)(ii) is heavily distorted compared to the transmitted signal Fig.~\ref{fig:Lines_and_rates}(a)(i) due to the slow rise and fall times of the atoms in this configuration. The video from this received signal, Fig.~\ref{fig:Lines_and_rates}(a)(iii), is very blurry with no color present. For a beam size of 400~$\mu$m (Fig.~\ref{fig:Lines_and_rates}(b)), we see that the received signal is less distorted, and the received video Fig.~\ref{fig:Lines_and_rates}(b)(iii) is clear but lacks color information. For a beam size of 200~$\mu$m (Fig.~\ref{fig:Lines_and_rates}(c)), we see that the received signal is less distorted, and the received video Fig.~\ref{fig:Lines_and_rates}(c)(iii) is slightly blury and contains color information. However, there are splotches where the color identification fails. For a beam size of 85~$\mu$m (Fig.~\ref{fig:Lines_and_rates}(c)), we see that the received signal is less distorted, and the received video Fig.~\ref{fig:Lines_and_rates}(c)(iii) is nearly as sharp as the direct image in Fig.~\ref{fig:NTCS_disc} (a) and has all the color information with correct saturation.

Demodulating the color information requires a higher bandwidth ($\approx3~$MHz) than the black and white intensity information. By optimizing the atomic response times according to Table~\ref{tab:rise_fall_times} with a probe beam FWHM = $85~\mu$m, we were able to achieve video reception that was not only sharp, but also captured color information.
To determine the data rate that was received, we calculate the effective bit rate of the transmitted signal.
The video camera outputs 480i video, interlacing 240 rows (scan lines) of 720 pixels at a field rate of 60~Hz to give a composite $480\times720$ pixels every 30~Hz.
If each pixel carries 24 bits of color information (8 bits for each of red, green, blue) and the nominal bitrate for 480i is 249~Mbps~$\approx60\frac{fields}{s}\times240\frac{lines}{field}\times720\frac{px}{line}\times24\frac{color~bits}{px}$, not including the information downtime during temporal alignment parts of the signal. 
Even though our Rydberg response rate and the detector bandwidth both fall well below this rate, we are able to recover enough phase information from the 3.58~MHz color carrier to display 480i color video using an analog-to-digital video converter. 

\begin{figure}
    \centering
    \includegraphics[width = \columnwidth]{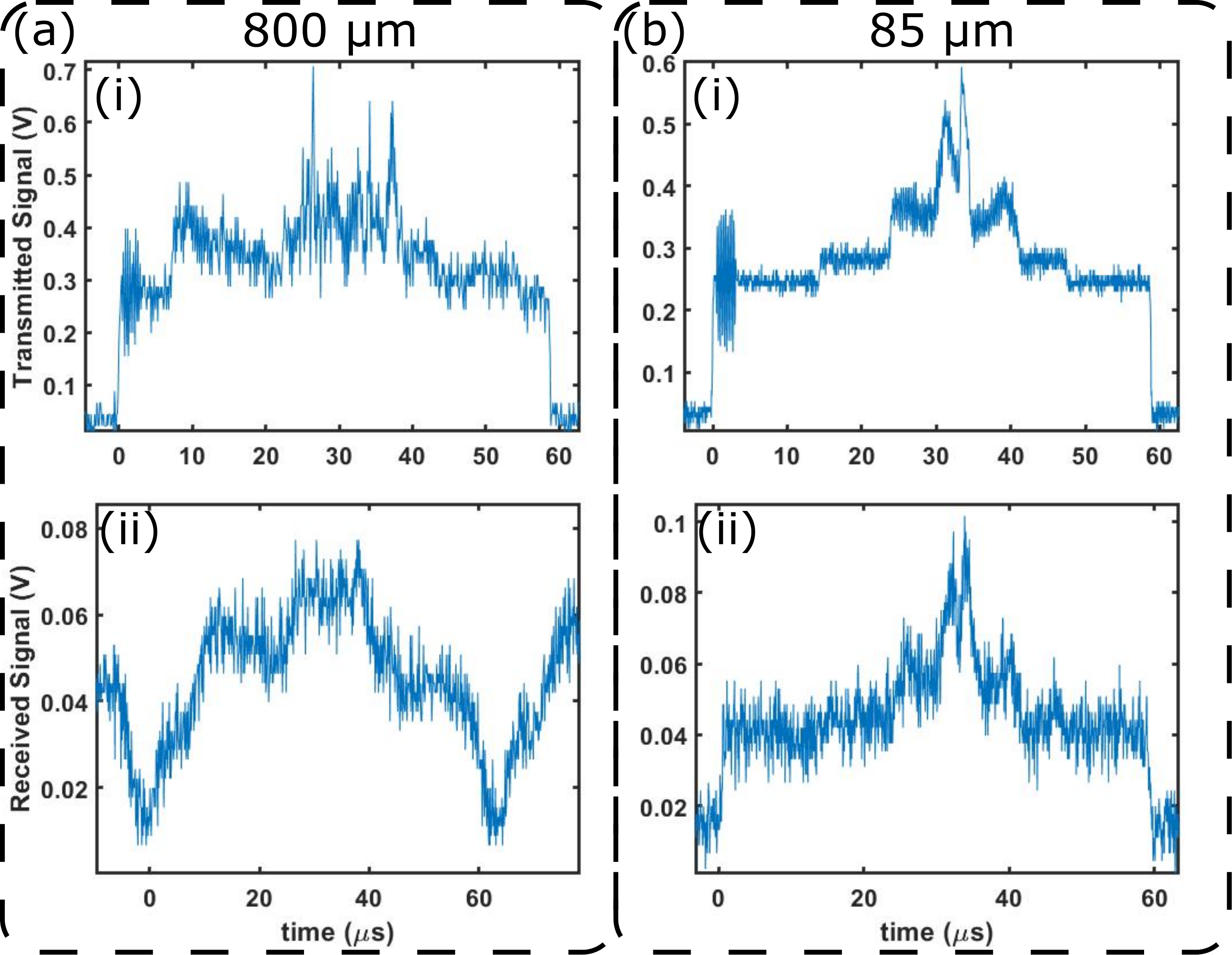}
    \caption{(a,b) Columns show sent and received data from game console for different beam sizes as labeled. (i) and (ii) are same as shown in Fig.~\ref{fig:Lines_and_rates}}
    \label{fig:game_reception}
\end{figure}

As a final example, we show data from streaming a game console. The video game console also outputs `standard definition' NTSC 480i video. Fig.~\ref{fig:game_reception} show the transmitted Fig.~\ref{fig:game_reception} (i) and received data Fig.~\ref{fig:game_reception} (ii) from a game console as the game is being played. Fig.~\ref{fig:game_reception}(a) is for a beam size of 800~$\mu$m and this resulted in a blurry image, Fig.~\ref{fig:game_reception} (c) is for a beam size of 85~$\mu$m and results is a clear color image. The stability of receiving the game signal is illustrates in the fact that the game could be played in real-time for several hours without losing the signal or its color clarity.

\section{Conclusion}

We have demonstrated the ability of the Rydberg atom receiver to receive live color video, using both a camera and a game console. The bandwidth was improved by optimizing the probe laser beam width, which tuned the average time the atoms remained in the interaction volume. We also show the dependence of EIT height, linewidth, rise time, and fall time on the probe beam width and Rabi frequency. EIT height and width both depend on Rabi frequency, but only the EIT height depends on beam width. The rise and fall times of the atomic response did not depend on the Rabi frequency, but show a proportional dependence on the beam width. This dependence is due to the transit time of the atoms through the cross section of the beam. Additionally, we used homoydne detection to overcome the limitation of our detector gain-bandwidth product by pre-amplifying the signal.
The bandwidth depends on the beam sizes and powers which ultimately determines the clarity of the reception and if color can be received. Finally, we achieved a data rate of 249~Mbps with this receiver.

\section*{Acknowledgements}
\vspace{-5mm}
\noindent This work was partially funded by the DARPA Quantum Apertures program and by the NIST-on-a-Chip (NOAC) Program.

\section*{AUTHOR DECLARATIONS}
\vspace{-3mm}
{\bf Conflict of Interest}\\
The authors have no conflicts to disclose.

\section*{Data Availability Statement}
\vspace{-5mm}\noindent Data is available upon request.

\end{document}